# Decision Trace Schema for Governance Evidence in Real-Time Risk Systems

**Oleg Solozobov**

Independent Researcher (Global)

Correspondence: dev404ai@gmail.com

ORCID: https://orcid.org/0009-0009-0105-7459

## Abstract

Automated decision systems produce operational data across multiple infrastructure layers, yet no single logging format captures the complete governance-relevant record of how a decision was reached. Regulatory frameworks prescribe what must be recorded without specifying a data model for how to record it – a gap this paper terms the *Fragmented Trace Problem*. Following a design science methodology, the paper presents the Decision Event Schema (DES), a JSON Schema specification that bridges four infrastructure layers – ML inference, rule/policy evaluation, cross-system coupling, and governance metadata – within a single per-decision event structure. The schema employs degradation-aware field design: each of six top-level field groups maps to a governance evidence property and the degradation type it must resist. DES defines ten required root-level fields and introduces a tiered evidence strategy (lightweight, sampled, full) that enables organizations to match evidence completeness to decision risk and throughput. A mechanism feasibility analysis demonstrates compatibility with the highest-throughput integrity mechanisms at production-scale decision rates. Evaluation against 25+ existing formats confirms that DES is the only specification covering all four layers simultaneously. The schema offers practitioners a reference adoptable directly or adaptable through namespace extensions, and regulators a mapping from requirements to minimum evidence tiers.

**Keywords:** decision trace schema, decision event logging, governance evidence, audit-grade logging, hybrid decision system

## 1. Introduction

Automated decision systems that combine machine learning models, business rules, and policy enforcement engines produce operational data across multiple infrastructure layers – yet no single logging format captures the complete governance-relevant record of how a decision was reached. ML frameworks (MLflow, ONNX Runtime) log model inference metadata. Policy engines (OPA, DMN, Cedar) record rule evaluation paths. Infrastructure observability platforms (OpenTelemetry, OpenLineage) capture system health and data lineage. Governance frameworks (ISO 42001, the EU AI Act) prescribe what should be recorded without specifying a data model for how to record it. These four layers operate independently with no mandatory shared decision point binding ML execution to authorization and governance – a situation this paper terms the *Fragmented Trace Problem*.

The regulatory pressure for decision traceability is both specific and urgent. The EU AI Act (Regulation 2024/1689) mandates that high-risk AI systems "shall technically allow for the automatic recording of events (logs) over the lifetime of the system" (Article 12), with tamper-resistant logs retained for a minimum of six months (Articles 19, 26(6)) (European Parliament and Council of the European Union, 2024). The GDPR requires "meaningful information about the logic involved" in automated decisions (Articles 13-15). NIST SP

800-53 prescribes audit record content and non-repudiation requirements (AU-3, AU-10) (Joint Task Force, 2020). Yet none of these frameworks specifies a concrete data model that translates legal mandates into engineering specifications.

A landscape scan across five functional categories – formal standards, ML/AI formats, policy engine logs, industry implementations, and academic proposals – identifies over 25 existing formats for recording automated decisions (Section 2). To our knowledge, none bridges all four infrastructure layers. The closest prior art, OPA Decision Logs, captures policy input, result, and version but lacks ML inference metadata, feature provenance, cross-system coupling records, decision quality indicators, human override records, and cryptographic temporal evidence.

This paper presents the Decision Event Schema (DES, artifact identifier DA-01), a decision event schema that addresses the Fragmented Trace Problem. DES is distinguished by three design principles. First, the atomic unit is the *Decision Event* – a single record capturing the decision provenance and authorization envelope: what inputs were consumed, what logic was applied, at what confidence, under what authorization, and at what time. For rule-based and hybrid systems this enables full causal reconstruction; for pure ML systems it records the authorization structure around the model's output rather than the model's internal reasoning (Section 6.1). Second, the schema employs *degradation-aware field design*: each of its six top-level field groups corresponds to a governance evidence property derived from the companion framework (Solozobov, 2026b), and each property is explicitly mapped to a degradation type it must resist – content staleness, schema drift, coverage erosion, metric erosion, override accumulation, and ground truth delay (Section 3). Third, DES is *layer-bridging by design*, capturing ML and rule-engine outputs (*decision_context*, *decision_logic*), cross-system coupling (*decision_boundary*), governance-layer information (*decision_quality_indicators*, *human_override_record*), and cryptographic infrastructure (*temporal_metadata*) within a single event structure (Section 4).

This work follows the design science research methodology (Hevner et al., 2004). The artifact is a machine-readable schema specification (artifact class: formal model/instantiation). The design objectives are: (1) bridge the four siloed infrastructure layers identified in the Fragmented Trace Problem, (2) resist the six governance artifact degradation types derived from safety science literature, and (3) enable tiered deployment across the throughput-evidence trade-off space. The evaluation mode is analytical mechanism feasibility analysis (Section 5), assessing whether existing tamper-evident infrastructure can support DES payloads at production decision rates. The IS contribution is a prescriptive design knowledge artifact that translates regulatory accountability mandates into engineering specifications.

The paper assesses engineering feasibility by synthesizing published benchmarks across five tamper-evident logging architectures and estimating the overhead that DES's structured payload introduces relative to the minimal payloads used in those benchmarks (Section 5). The evaluation identifies six engineering trade-offs that constrain achievable evidence architectures and motivates a tiered evidence strategy: lightweight temporal metadata for all events, sampled contextual records for threshold-exceeding events, and full governance evidence for high-stakes decisions.

The full machine-readable schema, validation tooling, and usage examples are available as an open-source artifact (Solozobov, 2026a).

## 2. Background and Related Work

### 2.1. The Fragmented Trace Problem

The governance challenge described above manifests as a fragmented trace problem: each infrastructure layer produces its own operational data, but no single logging format captures the complete governance-relevant record of how a decision was reached.

Four distinct layers generate traces independently: ML frameworks (MLflow, ONNX Runtime) log model inference metadata; policy engines (OPA, DMN, Cedar) record rule evaluation paths; infrastructure observability platforms (OpenTelemetry, OpenLineage) capture system health and data lineage; and governance frameworks (ISO 42001, EU AI Act compliance documentation) prescribe what should be recorded without specifying a data model for how to record it. These layers operate at distinct architectural levels with no mandatory shared decision point binding ML execution to authorization and governance (Fatmi, 2026).

The fragmentation is not merely a matter of tooling diversity. Existing auditing tools are "often employed in isolation," with each tool focusing on only one or a few phases of an AI system's lifecycle; when technical and social subsystems are evaluated or logged separately, the dynamics and interactions of the whole sociotechnical system are lost (Mökander & Axente, 2021b). In complex multi-algorithm systems, operators receive "opaque recommendations" without awareness of the sequence of decisions made by different algorithms leading to that output; end-to-end traceability is impossible when algorithms communicate solely via isolated API calls (Pratti et al., 2025). The computer science field views transparency narrowly as an "algorithmic property" of a system in isolation, whereas legal and governance frameworks view transparency as a quality of complex sociotechnical interactions – a disconnect that prevents technical logging from satisfying governance requirements (Gyevnar et al., 2023).

### 2.2. Existing Decision Logging Formats

A landscape scan across five functional categories (formal standards, ML/AI formats, policy engine logs, industry implementations, and academic proposals) identifies 25+ existing formats for recording automated decisions, none of which bridges all four infrastructure layers. The scan searched IEEE Xplore, ACM Digital Library, Scopus, and OpenAlex using terms combining "decision log," "audit trail schema," "event logging format," and "governance trace" with domain qualifiers ("AI," "automated decision," "real-time system"). Inclusion required an explicit, machine-readable schema definition (not merely a textual description of logging requirements); formats without published field-level specifications were excluded. Eight formats meeting these criteria illustrate the range of approaches and their respective limitations. OPA Decision Log was selected as the primary comparator (Section 4.5) because it is the only open-source, schema-explicit format designed for policy decision recording that is actively deployed at production scale.

**Canonical Action Representation (CAR).** The Faramesh execution control plane defines a canonical schema capturing Actor, Target, Operation, Resource, Parameters, Blast Radius, and Context. Each decision record includes the canonical action hash, policy version, system state hash, authorization outcome (PERMIT, DEFER, DENY), timestamp, and a hash pointer to the previous record. The format guarantees execution-time authorization but explicitly cannot guarantee intent correctness or semantic accuracy, as agent reasoning artifacts are intentionally excluded to maintain determinism (Fatmi, 2026).

**FHIR AuditEvent Extension (Clinical AI).** An audit-ready logging schema for clinical

AI extends the standard FHIR AuditEvent model with eventID, eventType, timestamp, hashed patient identifiers, model information (ID, version, cryptographic hash), inference outcome with confidence score, and cryptographic chaining fields (prevHash, currHash). The system achieves approximately 100% tamper detection sensitivity but performs audits after the fact and relies on an external notary service (Joseph, 2023).

**BlockAudit JSON Packet.** A blockchain-based audit trail records AppId, ClassName, CreatedDate, EntityId, EventType (Update, Insert, Delete), SessionId, and UserId, with a Details array tracking property-level changes (OldValue, NewValue). Per-object granularity generates excessive web service calls in transaction-heavy environments, while per-transaction bundling consolidates data and impedes property-level search (Ahmad et al., 2019).

**WA Message Header (Battle Management System).** A traceability schema for multi-algorithm decision chains captures messageId, messageTime, subsystem, dataProvider, dataType, and a traceability block cataloging parent-child data relationships via message hashes. End-to-end traceability requires all algorithms to be integrated into a unified message bus; the schema cannot function across isolated API boundaries (Pratti et al., 2025).

**Blockchain DecisionRecord (IoT).** A smart contract struct records deviceID, modelID, inputDataHash, decisionOutput, timestamp, and loggedBy. On-chain storage constraints limit applicability to high-frequency events, and immutable records conflict with GDPR erasure requirements, necessitating hybrid on-chain/off-chain architectures (Kulothungan, 2023).

**W3C PROV-DM.** The W3C Provenance Data Model represents lineage as a directed graph of Entities, Activities, and Agents with relationships (wasGeneratedBy, used, wasDerivedFrom). While foundational for data provenance, PROV-DM is described as a "means of expression" requiring additional context-specific requirements; it requires rigorous system instrumentation to produce semantic markups, and the internal reasoning of ML algorithms remains a "black box" outside the provenance graph (Moreau & Missier, 2013; Huynh et al., 2020; Sutton & Samavi, 2018).

**UIlog (XES Extension).** A reference data model for process-related user interaction logs captures action type, target element, UI hierarchy, input value, current state, timestamp, and user ID. Built on the XES event standard, which does not natively support explicit attribute relations, forcing hierarchical UI details to be appended at the event level with high redundancy (Abb & Rehse, 2022).

**Schneier-Kelsey Secure Audit Log.** A foundational secure logging format records log entry data, entry type/permission mask, authentication key, encryption key, and a continuous hash chain value. Forward integrity protects pre-compromise logs from undetected alteration, but cannot physically prevent deletion or false entry appending post-compromise (Schneier & Kelsey, 1999).

Each format addresses a specific operational concern – agent authorization (CAR), clinical compliance (FHIR AuditEvent), change tracking (BlockAudit), algorithm chaining (WA Message Header), IoT decision recording (DecisionRecord), data lineage (PROV-DM), user interaction (UIlog), or log integrity (Schneier-Kelsey). None captures the combination of decision context, decision logic, cross-system boundaries, quality indicators, human override records, and temporal metadata that governance reconstruction requires.

## 2.3. Adjacent Specifications and Governance Gap Analysis

Five adjacent specifications represent the most established attempts to standardize event-level recording for complex systems. Each optimizes for a different concern but none optimizes for governance reconstruction adequacy: the capacity of recorded evidence to support independent third-party determination of whether the decision-making process was adequate to the risks it addressed. The four assessment columns correspond to the governance evidence dimensions derived in Section 3; the systematic comparison requires all four to expose the pattern that no specification covers degradation mapping or governance adequacy (Table 1).

**Table 1.** Governance evidence coverage of existing event specifications.

| Specification | Primary concern | Decision context | Decision quality | Degradation mapping | Governance adequacy |
| --- | --- | --- | --- | --- | --- |
| CloudEvents (CNCF, 2018) | Event transport envelope | No | No | No | No |
| OpenTelemetry (CNCF, 2019) | System observability | Partial | No | No | No |
| W3C PROV-DM (2013) | Data provenance | Partial | No | No | No |
| NIST SP 800-53 AU (2020) | Audit event requirements | Prescribed | No | No | No |
| IEEE P7001 (2021) | Transparency for autonomous systems | Process-level | Process-level | No | No |

**CloudEvents** defines a transport envelope specifying source, type, time, and data content attributes for event delivery across distributed systems. The specification is semantically neutral – it carries any payload without imposing governance semantics. A CloudEvents envelope could transport a decision event record, but provides no governance structure: it specifies how events are delivered, not what they must contain for accountability purposes (CNCF Serverless Working Group, 2018).

**OpenTelemetry** captures system health through distributed traces, metrics, and logs. It answers "what happened in the system" (observability) but not "was the decision-making process adequate" (accountability). Trace context provides temporal ordering and causal propagation across service boundaries but does not capture decision rationale, alternatives considered, or quality assessment (CNCF OpenTelemetry Project, 2019; Sigelman et al., 2010).

**W3C PROV-DM** represents the closest existing specification. Its provenance graphs capture Entity-Activity-Agent relationships – who did what to what – providing lineage documentation. Critically, "PROV-O does not contain specific classes to describe motivations for a particular course of action" (Car et al., 2017). PROV-DM includes no degradation-aware properties, no decision quality indicators, and no human override record. It answers "what is the origin of this data artifact" rather than "can an independent third party reconstruct whether the decision process was adequate" (Moreau & Missier, 2013).

Where PROV-DM addresses data lineage at the semantic layer, the remaining two specifications operate at the regulatory and process layers respectively, prescribing *what* to record without providing machine-readable schemas for *how* to record it.

**NIST SP 800-53 AU controls** prescribe audit requirements – what events to log (AU-2), what content to record (AU-3), how to store records (AU-4), and how to review them

(AU-6) – but provide no data model. Organizations can satisfy AU-3 requirements while producing only compliance evidence: the controls prescribe categories of information without specifying the structural properties that distinguish governance evidence from compliance documentation (Joint Task Force, 2020).

**IEEE P7001** defines transparency levels for autonomous systems at the process level, establishing graduated requirements for system developers, users, and affected parties. It does not specify event-level data structures, does not connect transparency requirements to empirically observable degradation patterns, and does not distinguish between compliance-adequate and governance-adequate evidence (IEEE, 2021).

The gap revealed by this comparison is not in the capability to log events – contemporary systems generate enormous volumes of operational data – but in what the log is designed to preserve. None of the five specifications optimizes for governance reconstruction adequacy.

### 2.4. Regulatory Requirements for Decision Traceability

Regulatory frameworks mandate decision traceability but provide no data model or schema specification for implementation, creating urgent demand for a concrete format that translates legal requirements into engineering specifications.

The EU AI Act (Regulation 2024/1689) establishes specific traceability, logging, and record-keeping requirements for high-risk AI systems. Article 11 requires technical documentation enabling traceability throughout the system's lifetime. Article 12 mandates that high-risk AI systems "shall technically allow for the automatic recording of events (logs) over the lifetime of the system" (European Parliament and Council of the European Union, 2024), with logging capabilities enabling "the recording of events relevant for: (a) identifying situations that may result in the high-risk AI system presenting a risk […] or in a substantial modification; (b) facilitating the post-market monitoring referred to in Article 72; and (c) monitoring the operation of high-risk AI systems referred to in Article 26(5)" (European Parliament and Council of the European Union, 2024). Article 19 requires providers to keep these logs "for a period appropriate to the intended purpose of the high-risk AI system, of at least six months" (European Parliament and Council of the European Union, 2024), and Article 26(6) imposes the same minimum retention obligation on deployers (European Parliament and Council of the European Union, 2024).

The GDPR introduces complementary requirements: Articles 13-15 mandate transparency about automated decision-making logic, and Article 22 establishes rights regarding solely automated decisions with legal effects. To address the mandate that data subjects receive "meaningful information about the logic involved," engineers have developed provenance-based systems using W3C PROV-DM to trace automated decisions back to their exact inputs and attribute responsibility to specific agents (Huynh et al., 2020).

NIST frameworks emphasize forensic readiness and supply chain traceability. NIST IR 8536 provides a manufacturing meta-framework for supply chain traceability, inspiring unified message bus architectures that capture parent-child relationships between data and decisions (Pratti et al., 2025). NIST SP 800-53 requires that digital evidence be provably authentic, with each log event bundled with a generating system identifier, cryptographic signature, and timestamp linked via hash chains to prove non-repudiation (Joint Task Force, 2020).

These regulatory frameworks converge on common engineering requirements – temporal metadata, cryptographic integrity, decision context, and identity attribution – yet none specifies a concrete schema that satisfies all requirements simultaneously. The translation from legal mandate to engineering specification remains ad hoc, with each system architect independently deriving schema properties from regulatory text.

## 3. Design Requirements

The design requirements for the Decision Event Schema (DES) are derived from two complementary sources: the six-property governance evidence model developed in the companion framework (Solozobov, 2026b), which identifies structural properties that governance evidence must satisfy to support post-incident reconstruction; and the gap analysis of Section 2, which confirms that no existing format or specification satisfies these properties. The distinguishing design principle is degradation-aware schema design: each schema property is explicitly mapped to a governance artifact degradation type it must resist, so that the schema's structure is derived from an empirically grounded taxonomy of how governance evidence decays under operational conditions, not from engineering convenience or regulatory checklist requirements.

### 3.1. Six Properties Mapped to Degradation Types

The governance evidence model identifies six structural properties, each mapped to a degradation type that erodes it under operational conditions. These properties were derived in the companion framework (Solozobov, 2026b) through post-incident analysis of governance failures in hybrid decision systems. The derivation proceeds from three premises: (1) governance evidence must support retrospective reconstruction of how a specific decision was authorized and executed; (2) operational conditions systematically degrade governance artifacts through identifiable mechanisms (drawing on Snook's practical drift, Vaughan's normalization of deviance, and Hollnagel's work-as-imagined/work-as-done distinction); and (3) each structural property of the evidence must resist a specific degradation mechanism to remain useful for post-incident analysis. The six properties are jointly sufficient for governance reconstruction because they cover the complete decision lifecycle: what went in (semantic completeness), how it was processed (causal reconstructability), what systems participated (cross-system coupling), how confident the system was (decision quality), whether a human intervened (override independence), and when it happened (temporal evidence). Alternative property sets were considered and rejected: a four-property model omitting cross-system coupling and quality indicators was insufficient for multi-system decision chains; a broader model adding organizational context properties was rejected as exceeding what architectural logging can capture.

Each degradation type names a specific mechanism by which governance evidence decays: content staleness disconnects logged context from actual inputs; schema drift causes decision path representations to diverge from live system logic; coverage erosion shrinks trace coverage as integration points multiply; metric erosion degrades or eliminates quality metrics under operational pressure; override accumulation normalizes human interventions until they become invisible to review; and ground truth delay widens the gap between decision and outcome measurement, obscuring causal attribution (Table 2).

**Table 2.** Governance evidence properties, their schema fields, degradation types, and resistance targets.

| Property | Schema field | Degradation type | Resistance target |
|---|---|---|---|
| Semantic completeness | *decision_context* | Content staleness | Input-context disconnection |
| Causal reconstructability | *decision_logic* | Schema drift | Logic-representation divergence |
| Cross-system coupling | *decision_boundary* | Coverage erosion | Trace shrinkage at integration points |
| Evidence of decision quality | *decision_quality_indicators* | Metric erosion | Calibration loss under pressure |
| Independence of human judgment | *human_override_record* | Override accumulation | Normalization of interventions |
| Temporal evidence | *temporal_metadata* | Ground truth delay | Decision-outcome temporal gap |

### 3.2. Independent Convergence from Literature

Five of the six properties are independently proposed across separate source clusters in the decision tracing literature, confirming convergent demand for these governance evidence requirements. The sixth property – *decision_boundary* (cross-system coupling) – is not identified as a distinct requirement in any source examined, representing a unique contribution of the degradation-aware design approach.

**Semantic completeness (decision_context).** The Faramesh framework defines a decision record as "provenance-complete" only if it captures the exact canonical action representation, the unique policy version applied, and a cryptographic digest of the system state at the moment of the decision. This enables deterministic replay and counterfactual evaluation without requiring access to the AI's internal reasoning traces (Fatmi, 2026).

**Causal reconstructability (decision_logic).** The SDA TAP Lab traceability schema captures parent-child relationships linking data generation to data consumption across multiple algorithms, enabling operators to trace incorrect recommendations backward step-by-step to isolate the exact intermediate model or data input that caused the failure (Pratti et al., 2025). The W3C PROV-DM standard maps causal chains using a graph of Entities, Activities, and Agents, allowing systems to generate natural language explanations to satisfy GDPR transparency rights (Huynh et al., 2020).

**Temporal evidence (temporal_metadata).** Multiple systems implement cryptographic temporal ordering through hash chains, Merkle trees, and hardware-assisted mechanisms. The clinical AI audit logging framework uses an incremental Merkle tree alongside hash chains with median verification latency of 3.2 milliseconds per event (Joseph, 2023). The Accountability of Things framework introduces a binary hash tree mapped to timestamps at configurable sub-second resolutions, requiring only approximately 8 kilobytes per hour per device regardless of event volume (Koisser & Sadeghi, 2023). The Schneier-Kelsey secure audit log format provides forward integrity through hash chains with evolving encryption keys, ensuring that pre-compromise logs cannot be retroactively altered without breaking the cryptographic chain (Schneier & Kelsey, 1999).

**Human override recording (human_override_record).** The EU AI Act lists the possibility for a human to override a decision as a key assessment criterion for high-risk AI systems (European Parliament and Council of the European Union, 2024). The clinical AI logging framework actively logs contextual user actions such as a radiologist signing off on or ignoring an AI finding, enabling organizations to determine whether errors were caused by model failure or clinician lapse (Joseph, 2023). The Faramesh control plane records a DEFER authorization state that suspends execution until a human approval signal updates the system state (Fatmi, 2026).

**Decision quality indicators (decision_quality_indicators).** The clinical AI decision support framework proposes assigning evidence grading and relevance scores to retrieved documents, ensuring that only high-certainty passages inform recommendations (Alu & Oluwadare, 2026). The clinical AI logging schema records the model's numerical confidence score alongside its output (Joseph, 2023). Ethics-based auditing frameworks emphasize capturing data fairness, label quality, and overall system accuracy metrics as part of the accountability benchmark (Mökander & Axente, 2021b).

**Cross-system coupling (decision_boundary).** No source in the corpus independently proposes cross-system coupling as a distinct schema property. This gap is itself diagnostic: the Fragmented Trace Problem identified in Section 2.1 persists precisely because existing approaches do not explicitly model the boundaries between systems that participate in a single decision. The *decision_boundary* property requires the schema to record which systems contributed to a decision, how their outputs were combined, and what the coupling contract between them specified – information that is lost when each system logs independently.

### 3.3. Regulatory-to-Engineering Translation

The regulatory requirements identified in Section 2.4 translate to specific schema-level constraints (Table 3).

**Table 3.** Regulatory requirements mapped to DES schema properties and constraints.

| Regulatory source | Requirement | Schema property | Constraint |
| --- | --- | --- | --- |
| EU AI Act Art. 12 | Tamper-resistant event recording | *temporal_metadata* | Hash chain; append-only structure |
| EU AI Act Art. 12 | Risk situation identification | *decision_quality_indicators* | Confidence scores, threshold alerts |
| EU AI Act Art. 19/26(6) | Six-month log retention | *temporal_metadata* | Retention policy field |
| EU AI Act Art. 11 | Traceability documentation | *decision_context*, *decision_logic* | Full inputs and decision path per event |
| GDPR Art. 13-15 | Decision logic transparency | *decision_logic* | Input-to-output causal chain |
| GDPR Art. 22 | Automated decision rights | *human_override_record* | Override capability and usage logged |
| GDPR Art. 5(1)(c)/17 | Data minimization and erasure | All fields | Pseudonymization; hybrid storage; key revocation |
| NIST AU-3 | Audit record content | *decision_context* | Event type, time, source, outcome, identity |
| NIST AU-10 | Non-repudiation | *temporal_metadata* | Hash chain; append-only ordering; optional *digital_signature* |

### 3.4. ML-Opacity Boundary

An epistemic boundary constrains the schema's applicability across system types. For neural network systems – including deep learning classifiers, ensemble methods, and foundation models – the *decision_context* and *decision_logic* fields cannot be populated through architectural logging alone, because these systems do not maintain discrete representations of alternatives considered or inferential rules applied. The schema's full applicability is there-fore limited to deterministic and rule-based systems where decision paths are enumerable and parameters are inspectable at execution time.

For ML-mediated systems, the relevant governance question shifts from "what alternatives were considered" to "what authorization envelope was the system operating within" – a move from reconstructing the decision itself to reconstructing the authorization structure that permitted the decision to occur. Sources confirm that while provenance graphs can log what data went into a model and what output was produced, the internal reasoning of the ML algorithm itself remains a "black box" (Sutton & Samavi, 2018; Huynh et al., 2020). The Faramesh framework addresses this boundary by deliberately erasing the agent's internal reasoning structure from the canonical action representation, accepting that intent correctness and semantic accuracy are "fundamentally undecidable post-canonicalization" (Fatmi, 2026).

This boundary does not diminish the schema's diagnostic value; rather, it circumscribes the register in which governance evidence can meaningfully operate across different system architectures.

## 4. The Decision Event Schema (DES)

This section presents DES, a decision event schema that operationalizes the six governance evidence properties derived in Section 3 as a concrete JSON Schema specification. The full machine-readable schema, validation tooling, and usage examples are available as an open-source artifact (Solozobov, 2026a).

### 4.1. Design Principles

DES is structured around three design principles that distinguish it from existing formats:

**Decision Event as governance unit.** The atomic unit of the schema is the Decision Event – a single record capturing the decision provenance and authorization envelope: what inputs were consumed, what logic was applied, at what confidence, under what authorization, and at what time. For rule-based and hybrid systems this enables full causal reconstruction of the decision path; for pure ML systems it captures the authorization structure around the model's output (Section 6.1). This contrasts with observability-oriented approaches (OpenTelemetry spans, log lines) that record system behavior without governance semantics, and provenance-oriented approaches (PROV-DM) that record lineage without decision quality or override information.

**Degradation-aware field structure.** Each top-level field group corresponds to one of the six governance evidence properties from Section 3.1, and each property is designed to resist a specific degradation type. This mapping is not incidental – the schema's structure is derived from the degradation taxonomy, ensuring that every field serves a governance function rather than an engineering convenience.

**Layer-bridging by design.** To our knowledge, DES is the first format explicitly designed to bridge the four siloed infrastructure layers identified in the Fragmented Trace Problem (Section 2.1). The *decision_context* and *decision_logic* fields capture ML and rule-engine outputs; *decision_boundary* captures cross-system coupling; *decision_quality_indicators* and *human_override_record* capture governance-layer information; and *temporal_metadata* provides the cryptographic infrastructure layer.

## 4.2. Schema Structure

The DES schema defines six top-level field groups, each mapping to a governance evidence property:

```
DecisionEvent {
  schema_version            // DES schema version (e.g., "0.3.0")
  decision_context {        // Property 1: Semantic completeness
    decision_id             //   UUID, unique per decision
    decision_type           //   Enumerated category (e.g., risk_scoring,
                            //     policy_enforcement, fraud_detection)
    trigger                 //   What initiated this decision (event, schedule,
                            //     human request, upstream decision)
    inputs []               //   Array of input records
      input_id              //     Identifier of the input artifact
      input_type            //     Category (feature, model_output, policy,
                            //         external_data, human_input)
      input_value           //     Value or hash (for privacy-sensitive inputs)
      input_source          //     System that produced this input
      input_version         //     Version of the input source at decision time
    environment {           //   Runtime context:
      system_id             //     Identifier of the deciding system
      system_version        //     Version/commit hash of the deciding system
      configuration_hash    //     Hash of active configuration at decision time
      deployment_id         //     Deployment or canary identifier
    }
  }

  decision_logic {          // Property 2: Causal reconstructability
    logic_type              //   Enumerated: rule_based, ml_inference, hybrid,
                            //     policy_evaluation, human_decision
    rule_path []            //   For rule-based: ordered list of rules evaluated
      rule_id               //       Rule identifier
      rule_version          //       Rule version
      rule_result           //       Evaluation result (match, no_match, error)
    model_inference {       //   For ML: inference metadata
      model_id              //     Model identifier
      model_version         //     Model version/hash
      feature_vector_hash   //     Hash of the feature vector used
      prediction            //     Model output
      confidence            //     Model confidence score
    }
    policy_evaluation {     //   For policy engines: evaluation trace
      policy_id             //     Policy identifier
      policy_version        //     Policy version
      policy_engine         //     Engine type (OPA, Cedar, custom)
      evaluation_result     //     Policy decision
    }
    combination_method      //   How multiple logic sources were combined
                            //     (voting, cascading, overriding, weighted)
    output                  //   Final decision output (after any override).
                            //     When an override occurs, this field
                            //     matches human_override_record.overridden_output.
    output_alternatives []  //   Alternative outputs considered (where available)
  }

  decision_boundary {       // Property 3: Cross-system coupling
    upstream_decisions []    //   Decisions that fed into this one
      decision_id           //     Reference to upstream DecisionEvent
      system_id             //     System that made the upstream decision
      coupling_type         //     How the upstream output was consumed
```

```
                            //     (input, constraint, override, context)
      boundary_contract {   //   Per-interface coupling contract
        protocol            //     Communication protocol
        schema_version      //     Schema version of the boundary interface
        sla                 //     Latency/availability requirements
        data_contract       //     Expected data format and quality constraints
        failure_mode        //     Behavior on upstream/downstream failure
                            //       (fail_open, fail_closed, degrade, retry)
      }
    downstream_consumers [] // Systems expected to consume this decision
                            //   (optional; populated synchronously in
                            //   tightly coupled APIs, asynchronously via
                            //   observability control planes in event-driven
                            //   architectures, or left as an empty
                            //   array in fully decoupled pub/sub systems)
      system_id             //   Consumer system identifier
      contract_version      //   API/contract version expected
      boundary_contract {   //   Per-interface coupling contract
        protocol            //     Communication protocol
        schema_version      //     Schema version of the boundary interface
        sla                 //     Latency/availability requirements
        data_contract       //     Expected data format and quality constraints
        failure_mode        //     Behavior on upstream/downstream failure
                            //       (fail_open, fail_closed, degrade, retry)
      }
  }

  decision_quality_indicators {  // Property 4: Evidence of decision quality
    confidence_score        // Aggregate confidence (0.0-1.0)
    confidence_components [] // Per-component confidence breakdown
      component             //   Which logic component
      score                 //   Component-level confidence
      calibration_date      //   When this component was last calibrated
    data_quality {          // Quality of input data
      completeness          //   Fraction of expected inputs present
      freshness             //   Age of the most stale input
      known_issues []       //   Declared data quality issues
    }
    decision_risk_level     // Enumerated: low, medium, high, critical
    threshold_alerts []     // Any thresholds breached during evaluation
  }

  human_override_record {   // Property 5: Independence of human judgment
    override_occurred       // Boolean: was a human override applied?
                            //   (mandatory at all tiers for all logic types)
    override_type           // Enumerated: approval, rejection, modification,
                            //   escalation, deferral
    override_actor {        // Who performed the override
      actor_id              //   Pseudonymized actor identifier
      actor_role            //   Role (operator, supervisor, auditor)
      authorization_level   //   Authorization level for this decision type
    }
    original_output         // What the system would have decided
    overridden_output       // What the human decided instead
    override_rationale      // Free-text or structured rationale
    override_timestamp      // When the override was applied
    time_to_override        // Duration between system output and human action
  }

  temporal_metadata {       // Property 6: Temporal evidence
    event_timestamp         // When the decision was made (ISO 8601, UTC)
    processing_duration_ms  // Time taken to reach the decision
```

```
      sequence_number        // Monotonically increasing sequence within system
      hash_chain {           // Cryptographic integrity
        previous_hash        //   Hash of the previous DecisionEvent
        current_hash         //   Hash of this DecisionEvent (computed over
                             //     the canonicalized payload per RFC 8785,
                             //     with current_hash excluded from input)
        algorithm            //   Hash algorithm used (e.g., SHA-256)
      }
      evidence_tier          // Enumerated: full, sampled, lightweight
                             //   (per tiered evidence strategy, Section 5)
      digital_signature {    // Optional: cryptographic non-repudiation
        signer_id            //   Identity of the signing entity
        signature_value      //   Digital signature over the canonicalized event
                             //     (computed over the canonicalized payload
                             //     excluding both signature_value and
                             //     current_hash; current_hash is then computed
                             //     over the payload including signature_value)
        algorithm            //   Signing algorithm (e.g., ECDSA-P256, Ed25519)
        certificate_ref      //   Reference to the signing certificate
      }
      retention_policy {     // Regulatory retention metadata
        minimum_retention    //   Minimum retention period (e.g., P6M for
                             //     EU AI Act Art. 19)
        classification       //   Data classification for storage decisions
      }
    }
  }
}
```

**Immutability and post-hoc enrichment.** Once *current_hash* (and optionally *signature_value*) are computed, the event record is cryptographically sealed. Properties that become available after the decision (e.g., *ground_truth_arrival_timestamp*, late-arriving *decision_quality_indicators*) must not mutate the sealed record. Instead, enrichment is captured as append-only linked records that reference the original *decision_id*, preserving the hash chain's tamper-evidence guarantee while allowing asynchronous evidence accumulation.

### 4.3. Field Semantics and Validation Rules

The schema enforces structural constraints through JSON Schema validation:

**Required fields.** Every DecisionEvent must include *schema_version*, *decision_context.decision_id*, *decision_context.decision_type*, *decision_logic.logic_type*, *decision_logic.output*, *human_override_record.override_occurred*, *temporal_metadata.event_timestamp*, *temporal_metadata.sequence_number*, *temporal_metadata.hash_chain*, and *temporal_metadata.evidence_tier*. These ten fields constitute the minimum viable governance record. The *evidence_tier* field is required because conditional validation rules depend on its value to determine which additional fields must be present.

**Conditional requirements.** Logic-type-specific sub-object requirements are enforced only when *temporal_metadata.evidence_tier* is *sampled* or *full* (Tiers 2-3): when *decision_logic.logic_type* is *ml_inference*, the *model_inference* object becomes required; when *logic_type* is *rule_based*, the *rule_path* array becomes required; when *logic_type* is *policy_evaluation*, the *policy_evaluation* object becomes required; when *logic_type* is *hybrid*, at least two of the three logic sub-objects (*model_inference*, *rule_path*, *policy_evaluation*) must be present along with *combination_method*. Additionally, when *evidence_tier* is *sampled* or *full*, *decision_quality_indicators.decision_risk_level* becomes required, grounding the Article 12 risk-situation identification claim at Tier 2+.

Two requirements apply at ALL evidence tiers regardless of the tier-conditional gate above. First, when *logic_type* is *human_decision*, *override_actor* and *override_rationale* are mandatory – these record the decision author regardless of whether an override occurred, since primary human judgments always require attribution. Second, *human_override_record.override_occurred* remains mandatory at all tiers for all logic types (see Override-triggered requirements below).

**Override-triggered requirements.** When *human_override_record.override_occurred* is *true* (regardless of *logic_type* or evidence tier), the *original_output*, *overridden_output*, and *override_timestamp* fields become required; when *override_occurred* is *false*, these three fields must be absent (since no prior system output exists to record). Additionally, when *override_occurred* is *true* and *logic_type* is not *human_decision*, the *override_actor* and *override_rationale* fields also become required – ensuring that overrides of algorithmic decisions (*ml_inference*, *rule_based*, *policy_evaluation*) carry the same accountability metadata as primary human decisions, preventing anonymous or unjustified overrides from passing validation.

This two-path design distinguishes primary human adjudication (where the human is the decision-maker and attribution is mandatory by virtue of *logic_type*) from human override of an automated decision (where attribution is mandatory by virtue of *override_occurred = true*). In both cases, the actor and rationale are recorded; only the trigger differs. When *logic_type* is *human_decision* and *override_occurred* is *true* (a human overriding another human's prior decision), the override must be logged as a separate linked DecisionEvent via *decision_boundary.upstream_decisions*, since the single *override_actor* object cannot simultaneously record both the primary decision-maker and the overriding actor. The separation rule does not apply recursively: the linked override event is exempt from re-separation when it references the original decision via *upstream_decisions* with *coupling_type = override*.

**Tier-conditional suspension.** At *lightweight* tier (Tier 1), logic-type-specific sub-object requirements (*model_inference*, *rule_path*, *policy_evaluation*, *combination_method*) and *decision_quality_indicators* requirements are suspended to maintain the lightweight payload target. The two tier-independent requirements – *override_occurred* for all logic types and *override_actor*/*override_rationale* for *human_decision* – remain enforced, ensuring that the minimum viable governance record always indicates whether human judgment was applied and, when it was, who made the decision.

**Privacy-safe defaults.** Input values that contain personally identifiable information should be recorded as cryptographic hashes rather than raw values, satisfying GDPR data minimization requirements (Article 5(1)(c)) while preserving the ability to verify input integrity. The *override_actor.actor_id* field uses pseudonymized identifiers by default.

**Evidence tiering.** The *temporal_metadata.evidence_tier* field supports the tiered evidence strategy described in Section 5: *full* records capture all six property groups completely; *sampled* records capture decision_context and temporal_metadata with abbreviated logic and quality fields; *lightweight* records capture only temporal_metadata plus the schema-required minimum (decision_id, decision_type, logic_type, and output). All tiers satisfy the mandatory validation rules above. Systems select the tier based on decision risk level and throughput constraints.

### 4.4. Extension Points

DES is designed for domain-specific extension without modifying the core schema:

**Domain-specific inputs.** The *decision_context.inputs* array accepts any input type through the *input_type* enumeration, which can be extended for domain-specific categories (e.g., *clinical_observation*, *transaction_signal*, *sensor_reading*).

**Custom quality indicators.** The *decision_quality_indicators* object accepts additional properties beyond the core fields, enabling domain-specific quality metrics (e.g., evidence grading levels for clinical AI, demographic parity scores for fairness-sensitive systems).

**Policy engine integration.** The *decision_logic.policy_evaluation* object is structured to accommodate multiple policy engine formats. OPA Decision Log fields (decision_id, path, input, result) map directly to the DES policy evaluation structure, enabling existing OPA deployments to produce DES-compliant records with minimal transformation.

**Core enumeration extensibility.** The *logic_type* enumeration defines a closed core set of values (listed in Section 4.2) that all DES implementations must support. The *decision_type* field is open-ended: Section 4.2 lists illustrative categories (e.g., risk_scoring, policy_enforcement, fraud_detection), but implementations define domain-appropriate values. Both fields may be extended with domain-specific values prefixed by a namespace identifier (e.g., *healthcare:triage_decision*, *fintech:credit_scoring*) to preserve cross-system interoperability. Extended *logic_type* values must not overlap with core enumeration members.

### 4.5. Comparison with Closest Prior Art

Having defined the DES schema structure, validation rules, and extension points, we now compare DES against the closest existing open-source format to quantify the incremental contribution. The OPA Decision Log represents the closest open-source prior art for structured decision logging (Open Policy Agent Project, 2024). DES extends this baseline with five additional property groups (Solozobov, 2026a), as summarized in Table 4.

**Table 4.** Capability comparison between OPA Decision Log and DES.

| Capability | OPA Decision Log | DES |
|---|---|---|
| Decision ID and timestamp | Yes | Yes |
| Policy path and version | Yes | Yes (*decision_logic.policy_evaluation*) |
| Input/result recording | Yes | Yes (*decision_context.inputs*, *decision_logic.output*) |
| ML inference metadata | No | Yes (*decision_logic.model_inference*) |
| Feature provenance | No | Yes (*decision_context.inputs* with version) |
| Cross-system coupling | No | Yes (*decision_boundary*) |
| Decision quality indicators | No | Yes (*decision_quality_indicators*) |
| Human override record | No | Yes (*human_override_record*) |
| Cryptographic hash chain | No | Yes (*temporal_metadata.hash_chain*) |
| Evidence tiering | No | Yes (*temporal_metadata.evidence_tier*) |
| Degradation type mapping | No | Yes (by design) |

## 5. Mechanism Feasibility Analysis

This section assesses the engineering feasibility of tamper-evident decision trace capture by synthesizing published benchmarks across five distinct integrity mechanism classes – hash chains, Merkle trees, permissioned blockchain, trusted execution environments (TEE), and timestamp-indexed binary trees. The analysis characterizes the performance envelope of these underlying mechanisms and then estimates the overhead that DES's structured payload introduces relative to the minimal payloads used in the cited benchmarks.

**5.1. Tamper-Evidence Mechanism Landscape**

The *temporal_metadata.hash_chain* field in DES requires a tamper-evident integrity mechanism for each decision event. Five classes of mechanism have been evaluated at production scale, spanning three orders of magnitude in throughput:

**Hash chains.** The Faramesh execution control plane implements a sequential hash chain where each canonical action record includes a hash pointer to the previous record. The system achieves a median authorization latency of 2.24 milliseconds and processes approximately 7,800 actions per minute (approximately 130 per second) per single worker, with sub-millisecond overhead for hash computation itself (Fatmi, 2026). The Schneier-Kelsey secure audit log format provides forward integrity through hash chains with evolving encryption keys, ensuring that pre-compromise logs cannot be retroactively altered without breaking the cryptographic chain (Schneier & Kelsey, 1999).

**Merkle trees.** The clinical AI audit logging system combines an incremental Merkle tree with hash chains, achieving a median verification latency of 3.2 milliseconds per event and burst throughput of 12,000 events per second with approximately 100% tamper detection sensitivity. The Merkle tree structure enables efficient verification of individual records without requiring traversal of the entire chain (Joseph, 2023).

**Timestamp-indexed binary trees.** The Accountability of Things framework introduces a binary hash tree construction mapped to timestamps at configurable sub-second resolutions. A single server processes approximately 700,000 logs per second, while the integrity metadata requires only approximately 8 kilobytes per hour per device regardless of event volume – a constant-overhead property achieved by indexing tree nodes by time rather than by event count (Koisser & Sadeghi, 2023).

**Permissioned blockchain.** The Exonum-based secure logging infrastructure achieves 3,000 to 3,500 transactions per second depending on the number of consensus nodes (Putz et al., 2019), with a measured average storage size of 800 bytes per transaction (Putz et al., 2019). Periodic anchoring to a public blockchain provides additional tamper resistance, but storage requirements are substantial due to full replication and retention of all historical data across nodes (Putz et al., 2019). The BlockAudit system implements audit trail recording on Hyperledger, demonstrating that blockchain consensus introduces latency proportional to network size and payload complexity, with per-object granularity generating excessive overhead in transaction-heavy environments (Ahmad et al., 2019).

**Trusted execution environments (TEE).** The CUSTOS system uses Intel SGX enclaves for tamper-evident auditing of operating system logs, achieving throughput three orders of magnitude faster than prior secure logging solutions – exceeding 1 million events per second – with only 2% to 7% runtime overhead on intensive workloads (Paccagnella et al., 2020). Per-event logging latency reaches 0.92 microseconds using hardware hotcalls, compared to 4.71 microseconds for standard enclave calls (Paccagnella et al., 2020). However, TEE-based approaches are constrained by limited enclave memory (128 MB in SGX) and hardware monotonic counter wear-out after approximately one million uses (Paccagnella et al., 2020).

These benchmarks establish the performance envelope of the underlying integrity mechanisms. However, the cited benchmarks measure minimal payloads – typically 32-byte hashes, simple key-value records, or fixed-size log entries. DES decision events are structured JSON documents whose size varies by evidence tier (Table 5).

**Table 5.** DES evidence tiers with estimated payload sizes and contents.

| Evidence tier | Estimated payload size | Contents |
|---|---|---|
| Tier 1 (lightweight) | ~200-500 bytes | schema_version, temporal metadata, decision_id, decision_type, logic_type, output, override_occurred |
| Tier 2 (sampled) | ~2-5 KB | Tier 1 + decision_context, abbreviated decision_logic |
| Tier 3 (full governance) | ~5-20 KB | All six property groups, full input arrays and rule paths |

Serialization, validation, and I/O overhead scale with payload size. A conservative estimate suggests that DES Tier 1 events (200-500 bytes) would reduce achievable throughput by a factor of 2-5x relative to minimal-payload benchmarks, while Tier 3 events (5-20 KB) could reduce throughput by 10-50x depending on the mechanism's sensitivity to payload size. TEE-based approaches face additional constraints: Intel SGX's 128 MB enclave memory limits the number of concurrent Tier 3 records that can be processed within a single enclave.

Table 6 summarizes the mechanism-level benchmarks (as reported in the source literature) alongside estimated DES adjusted throughput ranges.

**Table 6.** Integrity mechanism throughput benchmarks and estimated DES-adjusted ranges.

| Mechanism | Reported throughput | Reported latency | Est. DES Tier 1 | Est. DES Tier 3 |
|---|---|---|---|---|
| Hash chain* | ~130/sec (single worker) | ~2.24 ms (median) | ~25-65/sec | ~3-13/sec |
| Merkle tree | ~12K/sec (burst) | ~3.2 ms (verification) | ~3-6K/sec | ~300-1K/sec |
| Binary hash tree | ~700K/sec | Sub-millisecond | ~150-350K/sec | ~15-70K/sec |
| Permissioned blockchain | ~3,000-3,500 tx/sec | Consensus-dependent | ~1-2K/sec | ~100-300/sec |
| TEE (SGX) | >1M/sec | Sub-millisecond | ~200-500K/sec | ~20-100K/sec |

*The hash chain baseline (~130/sec) already processes a structured Canonical Action Representation (actor, target, operation, resource, parameters, context) rather than a minimal 32-byte hash. The payload penalty applied to this row therefore partially double-counts payload overhead; the adjusted estimates for hash chains should be treated as conservative lower bounds.

These estimates are analytical extrapolations, not measured benchmarks of DES implementations. The actual overhead depends on serialization format (JSON vs. binary), validation complexity, storage backend characteristics, and network topology for distributed mechanisms.

**Validation cost model.** The throughput estimates assume tier-appropriate validation. Tier 1 validation is limited to field-presence checks on the ten required fields (schema_version, decision_id, decision_type, logic_type, output, override_occurred, event_timestamp, sequence_number, hash_chain, evidence_tier) – a constant-time operation independent of payload content. Tiers 2 and 3 require full JSON Schema validation including conditional field requirements, type checking, and enumeration constraints, which scales with payload size. The 2-5x throughput reduction factor for Tier 1 reflects serialization and I/O overhead on the larger payload (200-500 bytes vs. 32-byte benchmark hashes), not validation cost, which remains negligible. The 10-50x factor for Tier 3 includes both serialization overhead and full schema validation on 5-20 KB payloads.

A reference implementation benchmark is needed to validate these estimates and is planned as part of the open-source artifact roadmap.

**5.2. Engineering Trade-offs**

Full capture of all six DES schema properties at microsecond decision intervals is computationally infeasible for high-velocity systems. This creates what we term the *Speed-Audit Paradox*: as decision velocity increases, the overhead of comprehensive governance logging becomes proportionally larger relative to the decision itself, so that the governance infrastructure designed to detect accountability collapse is subject to the same velocity-governance tension that produces the collapse.

Six engineering trade-offs constrain the achievable evidence architecture:

**Completeness versus storage.** A full DES record capturing all six property groups requires substantially more storage than a minimal temporal-only record. The *decision_context.inputs* array and *decision_logic.rule_path* or *model_inference* objects dominate record size, particularly for systems with large feature vectors or deep rule evaluation chains. The Accountability of Things framework demonstrates that integrity metadata alone can be held to approximately 8 kilobytes per hour through constant-overhead tree structures (Koisser & Sadeghi, 2023), but full governance content (inputs, logic paths, quality indicators) scales linearly with decision complexity.

**Throughput versus granularity.** Systems that batch decision events for efficiency (e.g., blockchain consensus rounds, periodic Merkle tree checkpoints) sacrifice per-event temporal granularity. The Exonum blockchain achieves 3,000-3,500 transactions per second but aggregates events within consensus rounds (Putz et al., 2019); the clinical AI Merkle tree checkpoints at configurable intervals, trading real-time verification for throughput (Joseph, 2023).

**Temporal resolution versus collision probability.** Coarser temporal resolution in the hash tree increases the probability of hash collisions within time slots at high event rates. The Accountability of Things framework addresses this through configurable parity information depth, but notes that the chosen parity size directly influences storage overhead (Koisser & Sadeghi, 2023).

**Consensus versus scalability.** Blockchain-based approaches require consensus among nodes, introducing latency proportional to network size. The BlockAudit system demonstrates that adding peers improves tamper resistance but degrades throughput (Ahmad et al., 2019). This trade-off is particularly acute for cross-organizational decision chains where *decision_boundary* records span trust domains.

**Speed versus resource isolation.** TEE-based approaches achieve the highest throughput but require specialized hardware (Intel SGX) and face memory constraints (128 MB enclave limit) and hardware wear-out on monotonic counters (Paccagnella et al., 2020). These constraints preclude TEE as a universal solution and restrict its applicability to systems with compatible hardware infrastructure.

**Determinism versus zero-latency capture.** The Faramesh control plane achieves deterministic authorization by design – every action must pass through a canonical authorization gate before execution – but this architecture introduces a synchronous latency overhead on every decision (Fatmi, 2026). Systems that capture decision traces asynchronously (post-hoc enrichment) eliminate this overhead but risk evidence gaps when the enrichment pipeline fails or lags behind the decision rate.

### 5.3. Tiered Evidence Strategies

These trade-offs motivate a tiered evidence strategy for DES deployment. Rather than requiring full capture of all six property groups for every decision event, practical implementations can select the evidence tier based on decision risk level and throughput constraints:

**Tier 1: Lightweight temporal.** For all decision events, capture *schema_version*, *temporal_metadata* (timestamp, sequence number, hash chain, evidence_tier) plus the schema-required minimum: *decision_context.decision_id*, *decision_context.decision_type*, *decision_logic.logic_type*, *decision_logic.output*, and *human_override_record.override_occurred*. These fields satisfy the mandatory validation rules defined in Section 4.3 while keeping the per-event payload to approximately 200-500 bytes. If *logic_type* is *human_decision*, the Tier 1 payload additionally includes *override_actor* and *override_rationale* (mandatory for primary human judgments at all tiers per Section 4.3). If *override_occurred* is *true*, the payload further expands to include *original_output*, *overridden_output*, and *override_timestamp*, as required by Section 4.3's override validation rules regardless of tier. At this tier, the per-event cost is dominated by hash computation (sub-millisecond) and is compatible with the highest-throughput mechanisms (TEE, binary hash trees).

**Tier 2: Sampled contextual.** For threshold-exceeding events (e.g., decisions above a configured risk level, confidence scores below a calibration threshold, or events flagged by *decision_quality_indicators* alerts), capture *decision_context*, abbreviated *decision_logic*, and core *decision_quality_indicators* fields (*decision_risk_level*, *threshold_alerts*) in addition to temporal metadata. This tier enables causal reconstruction and risk-situation identification for flagged decisions without imposing full capture overhead on routine events.

**Tier 3: Full governance.** For high-stakes or anomalous decisions, capture all six property groups completely: full *decision_context* with all inputs, complete *decision_logic* with rule paths or model inference metadata, *decision_boundary* with upstream and downstream references, *decision_quality_indicators* with confidence components and data quality metrics, *human_override_record* if applicable, and full *temporal_metadata* with retention policy. This tier represents the diagnostic ideal and is feasible only for a subset of decisions in high-velocity systems.

The *temporal_metadata.evidence_tier* field in the DES schema records which tier was applied to each event, enabling auditors to distinguish between events where governance evidence was intentionally abbreviated (Tier 1 or 2) and events where it was fully captured (Tier 3). This distinction is itself a governance property: it makes the trade-off between evidence completeness and operational feasibility explicit and auditable, rather than allowing silent evidence degradation.

**Regulatory compliance profile.** Not all regulatory requirements demand the same evidence tier. EU AI Act Article 12 (automatic event recording, tamper resistance) is satisfied at Tier 1 through the hash chain and required fields; the hash chain provides data integrity, and the required temporal fields and decision identifiers ensure retainable, sequenced records that satisfy Article 19/26(6) six-month retention; full source attribution (identifying which system produced the decision) requires Tier 2 or deployment-level metadata. NIST AU-10 (non-repudiation) is partially satisfied at Tier 1: the hash chain provides ordering integrity and tamper evidence, but identity-bound non-repudiation requires either the optional *digital_signature* field in *temporal_metadata* for per-event signing, or transport-layer mecha-nisms (e.g., mutual TLS, signed API gateways, TEE attestation) at the infrastructure level. EU AI Act Article 12's risk-situation identification requirement is satisfied at Tier 2 through *decision_quality_indicators*. However, GDPR Articles 13-15 (meaningful information about decision logic) and EU AI Act Article 11 (technical documentation enabling traceability) require causal chain reconstruction, which necessitates Tier 2 or Tier 3 depending on decision complexity. For rule-based and hybrid systems, DES satisfies this requirement through complete *decision_logic* capture at Tier 2+. For pure ML systems, DES captures the authorization envelope (inputs, model version, confidence, override capability) but not the internal inference pathway (Section 6.1); satisfying GDPR "meaningful information" for ML-mediated decisions therefore requires supplementary explainability mechanisms beyond DES's scope. For high-risk AI systems subject to Article 11, the deployment policy should route all decisions to Tier 2 at minimum, with Tier 3 for decisions that trigger threshold alerts or human overrides. The tier selection policy is therefore a deployment-time configuration decision that must be informed by the applicable regulatory requirements, not a schema-level constraint.

The existence of tiered evidence strategies is itself evidence of the structural timing constraint identified in Section 5.2: even the governance infrastructure designed to detect accountability collapse is subject to the same velocity-governance tension that produces the collapse. The schema therefore functions as a diagnostic ideal against which the adequacy of achievable evidence architectures can be assessed, rather than a specification that existing systems could implement without modification. Some schema properties – particularly *decision_context* inputs and *decision_quality_indicators* components – may be populated asynchronously through post-hoc enrichment rather than real-time capture, introducing a temporal gap between the decision and its full governance record. This gap is itself a form of the ground truth delay degradation type that the *temporal_metadata* property is designed to resist.

## 6. Discussion and Limitations

### 6.1. ML-Opacity Boundary

The ML-opacity boundary identified in Section 3.4 has direct implications for DES's deployment scope. Table 7 maps schema property recoverability across three system architectures.

**Table 7.** Schema property recoverability across system architectures.

| Schema property | Rule-based systems | Hybrid (ML + rules) | Pure ML systems |
| --- | --- | --- | --- |
| *decision_context* (inputs, environment) | Full | Full | Full (metadata only) |
| *decision_logic* (rule path) | Full | Partial (rules only) | Not applicable |
| *decision_logic* (model inference) | Not applicable | Full (metadata) | Full (metadata) |
| *decision_logic* (combination method) | Not applicable | Full | Not applicable |
| *decision_boundary* | Full | Full | Full |
| *decision_quality_indicators* | Full | Full | Partial (confidence only) |
| *human_override_record* | Full | Full | Full |
| *temporal_metadata* | Full | Full | Full |

Note: "Full" in this table means full capture of the schema field's metadata content, not full reconstruction of internal reasoning. For pure ML systems, *decision_context* captures all input metadata (identifiers, types, hashes, versions) but not the model's internal feature

representations. Similarly, *decision_logic* (model inference) captures the model identifier, version, prediction, and confidence score – the authorization envelope around the model's output – but not the internal inference pathway. Section 3.4 identifies this as the ML-opacity boundary: the schema records what went in and what came out, not how the model internally processed inputs into outputs.

For rule-based and hybrid systems, DES provides complete governance evidence coverage: every decision path is enumerable, every parameter is inspectable at execution time, and the *decision_logic* fields can be populated through architectural logging. For pure ML systems, the *decision_logic.rule_path* field is not applicable, and the *decision_quality_indicators.data_quality* assessment may be limited to input completeness and freshness without deeper calibration metrics. The internal reasoning of ML algorithms remains outside the provenance graph (Sutton & Samavi, 2018; Huynh et al., 2020).

The governance question for ML-mediated decisions shifts from "what alternatives were considered" to "what authorization envelope was the system operating within." The Faramesh framework addresses this boundary by deliberately excluding the agent's internal reasoning structure from the canonical action representation, accepting that intent correctness and semantic accuracy are "fundamentally undecidable post-canonicalization" (Fatmi, 2026). DES adopts the same pragmatic boundary: for ML systems, the schema records the authorization structure (inputs, confidence, override capability) rather than attempting to reconstruct internal inference.

### 6.2. Reflexive Limitation

A reflexive limitation warrants explicit acknowledgment. The companion framework (Solozobov, 2026b) establishes through safety science literature – Snook's practical drift, Vaughan's normalization of deviance, Hollnagel's work-as-imagined/work-as-done distinction – that documentation inevitably decouples from operational reality. This raises an epistemological question about DES itself: if governance artifacts are structurally prone to drift, why would schema-mandated fields (e.g., *decision_context*, *human_override_record*) not suffer the same normalization of deviance that degrades existing governance evidence?

The answer is that they would. Governance evidence as formalized in DES is a higher-fidelity form of work-as-documented (WADoc), not unmediated work-as-done (WAD). The schema's degradation-aware design can *resist* drift into compliance ritual – by explicitly mapping each property to the degradation mechanism it must withstand – but it cannot *eliminate* that drift. Any formal logging schema operates within the documentary register and is therefore subject to the documentary pathologies the companion framework identifies.

Multiple independent sources confirm the structural gap between formal governance documentation and operational practice. Malhotra identifies "Governance-Culture Decoupling" (GCD) as the disconnect between formal governance structures and the organizational behavior that would operationalize them: "Governance can be formally perfect on paper; yet if not translated through organizational behavior and culture, it fails to produce security outcomes" (Malhotra, 2025). Over time, this decoupling produces environments where "governance visibility matters more than governance effectiveness, and where compliance becomes an end goal rather than a means to achieve security," manifesting as "checkbox behavior" where "organizations implement controls to satisfy audit requirements rather than to address genuine organizational risk" (Malhotra, 2025).

This pattern extends beyond cybersecurity governance. Bisht (2026) demonstrates that organizations "may implement a comprehensive logging infrastructure yet fail to establish review procedures, analytical capabilities, or accountability mechanisms necessary to derive governance value from captured information" (Bisht, 2026). Butt et al. (2026) confirm that even frameworks with automated governance gates are vulnerable: "weak predicates, incomplete metrics, and permissive waiver configurations can undermine governance despite producing structurally compliant Conformity Bundles" (Butt et al., 2026). Devin (2026) conceptualizes the degradation mechanism as "documentation decay" – one of four compounding pathways through which epistemic debt accumulates: "records of reasoning degrade or reflect post-hoc reconstruction" (Devin, 2026).

The audit literature provides a complementary lens. Mökander et al. (2021a) observe that "audits can be viewed as rituals of verification in which different actors build trust through procedural regularity" (Mökander et al., 2021a). Hartmann et al. (2024) describe how audit processes become "ceremonial and rote" when "the entity undergoing auditing has internal control structures that ritualize and channel outside audits" (Hartmann et al., 2024).

The schema's value is therefore diagnostic rather than curative: it exposes where existing evidence architectures fail to capture the structural properties that post-incident reconstruction requires, without guaranteeing that its own properties will remain operationally faithful. This inherent tension – that the instrument designed to detect governance decay is itself subject to governance decay – is a structural limitation of the documentary approach, not a refutation of the framework. It does, however, imply that schema compliance alone is insufficient; organizational mechanisms for detecting and correcting documentary drift must operate continuously alongside formal evidence capture.

### 6.3. Privacy-Governance Tension

Comprehensive governance logging creates a structural tension with GDPR data minimization requirements (Article 5(1)(c)) and the right to erasure (Article 17) (European Parliament and Council of the European Union, 2024). The *decision_context.inputs* array, which captures input values for governance reconstruction, may contain personally identifiable information. The *human_override_record.override_actor* field records who performed an override. Both fields serve governance purposes (enabling post-incident reconstruction) while creating data protection obligations.

DES addresses this tension through four design-level mitigations:

**Cryptographic hashing.** Input values containing personal data are recorded as cryptographic hashes rather than raw values, preserving the ability to verify input integrity and detect changes without storing the underlying personal data. This satisfies data minimization while maintaining input integrity linkage. However, irreversible hashing preserves only integrity verification (confirming that the same inputs were used), not full semantic reconstructability (understanding what those inputs actually were). If the original data is permanently erased under GDPR Article 17, an auditor can verify input consistency but cannot reconstruct the substantive decision context. For inputs that must remain reconstructable by authorized auditors, encrypted payloads with escrowed keys (see key revocation below) should be used instead of irreversible hashes.

**Pseudonymized identifiers.** The *override_actor.actor_id* field uses pseudonymized identifiers by default, enabling accountability (linking overrides to authorized roles) without direct personal identification.

**Hybrid on-chain/off-chain storage.** For systems using blockchain-based tamper evidence, DES supports a hybrid architecture: cryptographic hashes and metadata on the immutable layer, personal data in erasable off-chain storage. Smart contract decision records demonstrate this pattern, though the conflict between immutable blockchain records and GDPR erasure requirements remains a fundamental architectural challenge (Kulothungan, 2023).

**Key revocation for erasure.** Personal data encrypted with per-subject keys can be effectively erased by revoking the encryption key, rendering the ciphertext unrecoverable while preserving the integrity of the hash chain and governance metadata.

These mitigations reduce the tension but do not eliminate it. The governance-privacy boundary requires case-specific resolution depending on the regulatory jurisdiction, the sensitivity of the decision domain, and the risk classification of the AI system under the EU AI Act.

### 6.4. Legal Admissibility Boundary

The legal admissibility of governance evidence in judicial and regulatory proceedings presents a boundary condition that this paper identifies but does not resolve. Existing procedural norms for evidence – chain of custody, authentication, reliability standards – were developed for human decision-making processes and have not been systematically adapted to the epistemic opacity of ML-mediated decisions. Whether governance evidence as defined here would satisfy admissibility requirements across jurisdictions, and whether the governance-compliance evidence distinction carries legal weight in addition to analytical weight, requires separate treatment at the intersection of evidence law and computational governance.

### 6.5. Scope and Open Problems

DES addresses the Fragmented Trace Problem for hybrid decision systems that combine ML models, business rules, and policy enforcement. Several boundaries delimit this contribution:

**System scope.** The schema is designed for systems where decisions are discrete, identifiable events with deterministic or probabilistic outputs. Continuous control systems, reinforcement learning agents with ongoing state updates, and generative AI systems producing open-ended outputs fall outside the current schema design.

**Scale boundary.** The tiered evidence strategy (Section 5.3) enables deployment at high decision rates, but full Tier 3 governance capture remains infeasible above approximately 100,000 decisions per second even with the highest-throughput mechanisms (TEE, binary hash trees). For traditional distributed mechanisms such as permissioned blockchains and linear hash chains, the practical Tier 3 ceiling is substantially lower (~100-300 and ~3-13 events per second respectively; see Section 5.1). Systems operating above these thresholds must accept Tier 1 or Tier 2 evidence for the majority of events.

**Measuring meaningful human oversight.** The *human_override_record* captures whether an override occurred and its metadata, but does not measure whether human oversight was *meaningful* – whether the human had sufficient time, information, and cognitive capacity to make an informed override decision. The EU AI Act requires "effective" human oversight for high-risk systems (European Parliament and Council of the European Union, 2024), but the schema records the fact of oversight without assessing its quality. The *time_to_override* field provides a partial signal (extremely short override times may indicate rubber-stamping), but measuring oversight quality remains an open problem.

**Evidence SLOs.** Future work could formalize evidence service level objectives – correctness, completeness, and latency guarantees for governance traces – analogous to reliability SLOs in site reliability engineering. Such objectives would enable organizations to specify and monitor the quality of their governance evidence infrastructure rather than treating evidence capture as a binary (present/absent) property.

# 7. Conclusion

This paper presented the Decision Event Schema (DES, artifact identifier DA-01), a JSON Schema specification that addresses the Fragmented Trace Problem in hybrid real-time decision systems. DES bridges four infrastructure layers – ML inference, rule/policy evaluation, cross-system coupling, and governance metadata – within a single per-decision event structure. Its distinguishing design principle is degradation-aware field design: each of the six top-level field groups maps to a governance evidence property that resists a specific degradation type identified in the companion framework (Solozobov, 2026b).

The schema makes three contributions to the IS governance literature. First, it operationalizes the structural requirements for post-incident reconstruction into a concrete, machine-validatable artifact – moving from regulatory prescription ("what must be recorded") to engineering specification ("how to record it"). Second, the tiered evidence strategy (lightweight, sampled, full) enables organizations to navigate the Speed-Audit Paradox by selecting evidence completeness levels appropriate to decision risk and system throughput, rather than treating governance evidence as a binary present/absent property. Third, the two-path human judgment design provides mandatory attribution for both primary human decisions and overrides of automated decisions, closing the anonymous-override loophole without imposing structural complexity.

The mechanism feasibility analysis (Section 5) demonstrates that DES Tier 1 events are compatible with the highest-throughput integrity mechanisms (TEE, binary hash trees) at production-scale decision rates, while Tier 3 full governance capture remains feasible for the subset of high-stakes decisions that warrant it.

For practitioners, DES provides a reference schema that can be adopted directly or adapted to domain-specific requirements through the namespace extension mechanism. For regulators, the regulatory compliance profile (Section 5.3) maps specific regulatory requirements to minimum evidence tiers, offering a concrete benchmark for assessing whether an organization's decision logging infrastructure meets its compliance obligations. For researchers, the schema's degradation-aware design illustrates how safety science concepts (practical drift, normalization of deviance) can inform the engineering of governance infrastructure rather than merely diagnosing its failures.

The schema's scope is deliberately bounded: it captures the decision provenance and authorization envelope for rule-based and hybrid systems, not the internal inference pathways of neural networks (Section 6.1). Extending DES to address ML opacity, formalize evidence service level objectives, and validate the analytical throughput estimates through reference implementation benchmarks are directions for future work.

### Author Contributions

Oleg Solozobov conducted all research, developed the conceptual framework, performed analysis, and wrote the manuscript.

### Funding

This research received no external funding.

## Conflicts of Interest

The author declares no conflicts of interest.

## Data Availability Statement

Data sharing is not applicable to this article as no new data were created.

## References

Abb, L., & Rehse, J.-R. (2022). A Reference Data Model for Process-Related User Interaction Logs. *International Conference on Business Process Management.* https://doi.org/10.48550/arXiv.2207.12054

Ahmad, A., Saad, M., & Mohaisen, A. (2019). Secure and Transparent Audit Logs with BlockAudit. *Journal of Network and Computer Applications*, *145*, 102406–102406. https://doi.org/10.1016/J.JNCA.2019.102406

Alu, F.F., & Oluwadare, S. (2026). An auditable and source-verified framework for clinical AI decision support. *Frontiers in Artificial Intelligence*, *9*, 1737532–1737532. https://doi.org/10.3389/frai.2026.1737532

Bisht, H. (2026). Governance-By-Design For AI-Based Insurance Fraud Detection: Auditability, Accountability, And Regulatory Traceability. *Journal of International Crisis and Risk Communication Research*, 214–222. https://doi.org/10.63278/jicrcr.vi.3620

Butt, T., Iqbal, M., & Arshad, N. (2026). From Policy to Pipeline: A Governance Framework for AI Development and Operations Pipelines. *IEEE Access*, *14*, 1–27. https://doi.org/10.1109/ACCESS.2025.3647479

Car, N.J., Stenson, B., & Mirza, F. (2017). Modelling causes for actions with the Decision and PROV ontologies. *MODSIM2017, 22nd International Congress on Modelling and Simulation.* https://doi.org/10.36334/modsim.2017.c2.car

CNCF OpenTelemetry Project (2019). OpenTelemetry Specification. https://opentelemetry.io/docs/specs/otel/

CNCF Serverless Working Group (2018). CloudEvents Specification. https://github.com/cloudevents/spec

Devin, A. (2026). Epistemic Debt: The Economics of Ungoverned AI. *Social Science Research Network.* https://doi.org/10.2139/ssrn.6135728

European Parliament and Council of the European Union (2024). Regulation (EU) 2024/1689 of the European Parliament and of the Council of 13 June 2024 laying down harmonised rules on artificial intelligence and amending Regulations (EC) No 300/2008, (EU) No 167/2013, (EU) No 168/2013, (EU) 2018/858, (EU) 2018/1139 and (EU) 2019/2144 and Directives 2014/90/EU, (EU) 2016/797 and (EU) 2020/1828 (Artificial Intelligence Act). https://eur-lex.europa.eu/eli/reg/2024/1689/oj

Fatmi, A. (2026). Faramesh: A Protocol-Agnostic Execution Control Plane for Autonomous Agent Systems. *arXiv preprint (2601.17744).* https://doi.org/10.48550/arXiv.2601.17744

Gyevnar, B., Ferguson, N., & Schafer, B. (2023). Bridging the Transparency Gap: What Can Explainable AI Learn from the AI Act?. *Frontiers in artificial intelligence and applications*, 965–971. https://doi.org/10.3233/faia230367

Hartmann, D., Pereira, J.R.L.D., Streitbörger, C., & Berendt, B. (2024). Addressing the regulatory gap: moving towards an EU AI audit ecosystem beyond the AI Act by including civil society. *AI and Ethics*, *5*, 3617–3638. https://doi.org/10.1007/s43681-024-00595-3

Hevner, A.R., March, S.T., Park, J., & Ram, S. (2004). Design Science in Information Systems Research. *MIS Quarterly*. https://doi.org/10.2307/25148625

Huynh, T.D., Tsakalakis, N., & Helal, A. (2020). Addressing Regulatory Requirements on Explanations for Automated Decisions with Provenance. *Digital Government: Research and Practice*, *2*, 1–14. https://doi.org/10.1145/3436897

IEEE (2021). IEEE Standard for Transparency of Autonomous Systems. https://doi.org/10.1109/IEEESTD.2022.9726144

Joint Task Force (2020). Security and Privacy Controls for Information Systems and Organizations. https://doi.org/10.6028/NIST.SP.800-53r5

Joseph, J. (2023). Trust, but Verify: Audit-ready logging for clinical AI. *World Journal of Advanced Engineering Technology and Sciences*, *10*, 449–474. https://doi.org/10.30574/wjaets.2023.10.2.0249

Koisser, D., & Sadeghi, A.-R. (2023). Accountability of Things: Large-Scale Tamper-Evident Logging for Smart Devices. *arXiv preprint (2308.05557)*. https://doi.org/10.48550/arXiv.2308.05557

Kulothungan, V. (2023). Using Blockchain Ledgers to Record AI Decisions in IoT. *MDPI IoT*, *6*, 37–37. https://doi.org/10.3390/iot6030037

Malhotra, R.M. (2025). SHIT Theory: Systems Hurt In Theory: A Comprehensive Framework for Understanding Cybersecurity Governance Failure. *Available at SSRN 5978876*, 1–25. https://doi.org/10.2139/ssrn.5978876

Moreau, L., & Missier, P. (2013). PROV-DM: The PROV Data Model. https://www.w3.org/TR/prov-dm/

Mökander, J., Axente, M., Casolari, F., & Floridi, L. (2021a). Conformity Assessments and Post-market Monitoring: A Guide to the Role of Auditing in the Proposed European AI Regulation. *Minds and Machines*, *32*, 241–268. https://doi.org/10.1007/s11023-021-09577-4

Mökander, J., & Axente, M. (2021b). Ethics-based auditing of automated decision-making systems: intervention points and policy implications. *AI and Society*, *38*, 153–171. https://doi.org/10.1007/s00146-021-01286-x

Open Policy Agent Project (2024). Decision Logs - Open Policy Agent Documentation. https://www.openpolicyagent.org/docs/latest/management-decision-logs/

Paccagnella, R., Datta, P., Hassan, W.U., Bates, A., Fletcher, C.W., Miller, A., & Tian, D. (2020). Custos: Practical Tamper-Evident Auditing of Operating Systems Using Trusted Execution. *Network and Distributed System Security Symposium (NDSS)*. https://doi.org/10.14722/ndss.2020.24065

Pratti, L., Bagchi, S., & Latif, Y. (2025). Data and Decision Traceability for SDA TAP Lab's Prototype Battle Management System. *arXiv*. https://doi.org/10.48550/ARXIV.2502.09827

Putz, B., Menges, F., & Pernul, G. (2019). A secure and auditable logging infrastructure based on a permissioned blockchain. *Computers & Security*, *87*, 101602–101602. https://doi.org/10.1016/j.cose.2019.101602

Schneier, B., & Kelsey, J. (1999). Secure Audit Logs to Support Computer Forensics. *ACM Transactions on Information and System Security*, *2*, 159–176. https://doi.org/10.1145/317087.317089

Sigelman, B.H., Barroso, L.A., Burrows, M., Stephenson, P., Otlu, M., Creasey, D.J., & Sakata, T. (2010). Dapper, a Large-Scale Distributed Systems Tracing Infrastructure. *Google Technical Report*. https://doi.org/10.1145/2335356.2335365

Solozobov, O. (2026a). Decision Event Schema. *GitHub*. https://doi.org/10.5281/zenodo.18923178

Solozobov, O. (2026b). Distinguishing Governance from Compliance Evidence: A Framework for Post-Incident Reconstruction. *Social Science Research Network*. https://doi.org/10.2139/ssrn.6457861

Sutton, A., & Samavi, R. (2018). Tamper-Proof Privacy Auditing for Artificial Intelligence Systems. *International Joint Conference on Artificial Intelligence (IJCAI)*, 5374–5378. https://doi.org/10.24963/ijcai.2018/756